\documentclass[dvips]{article}
\usepackage{subfigure}
\usepackage{amsmath}
\usepackage{amsmath,amsfonts,amssymb,amsopn}
\usepackage{icrctc07}

\newcommand{\be}{\begin{equation}}
\newcommand{\ee}{\end{equation}}
\newcommand{\ben}{\begin{eqnarray}}
\newcommand{\een}{\end{eqnarray}}
\newcommand{\bc}{\begin{center}}
\newcommand{\ec}{\end{center}}

\title{Geomagnetic Field Effects on the Imaging Air Shower Cherenkov
  Technique}
\authors{S.C. Commichau$^{1}$, A. Biland$^{1}$, D. Kranich$^{1}$, R. de los
Reyes$^{2}$, A. Moralejo$^{3}$, D. Sobczy\'nska$^{4}$ on behalf of the MAGIC
collaboration$^{*}$}

\shorttitle{GF Effects on the Imaging Air Shower Cherenkov
  Technique}

\shortauthors{S.C. Commichau et al.}

\afiliations{
$^1$ETH Zurich, Institute for Particle Physics, 8093 Zurich, Switzerland\\
$^2$Universidad Complutense, 28040 Madrid, Spain\\ 
$^3$Institut de F\'isica d'Altes Energies, Edifici Cn., 08193
Bellaterra (Barcelona), Spain\\ 
$^4$Division of Experimental Physics, University of Lodz, 90236, Poland\\ 
$^*$See \texttt{http://wwwmagic.mppmu.mpg.de/collaboration/members/}}

\email{sebastian.commichau@phys.ethz.ch}

\abstract{Imaging Air Cherenkov Telescopes (IACTs) detect the Cherenkov light flashes
of Extended Air Showers (EAS) triggered by very high energy (VHE) $\gamma$-rays impinging on the Earth's
atmosphere. Due to the overwhelming background from hadron induced
EAS, the discrimination of the rare $\gamma$-like events is rather difficult,
in particular at energies below 100\,GeV. The influence of the Geomagnetic
Field (GF) on the EAS development can further complicate this discrimination
and, in addition, also systematically affect the $\gamma$ efficiency and energy
resolution of an IACT. Here we present the results from dedicated Monte Carlo
(MC) simulations for the MAGIC telescope site. Additionally we show that
measurements of sub-TeV $\gamma$-rays from the Crab nebula are affected even
for a low GF strength of $|\vec{B}_{\perp}| < 30\,\mu\text{T}$.}

\begin{document}

\maketitle
\section{Introduction}

\begin{figure}[!h] \centering
  \includegraphics[scale=.33]{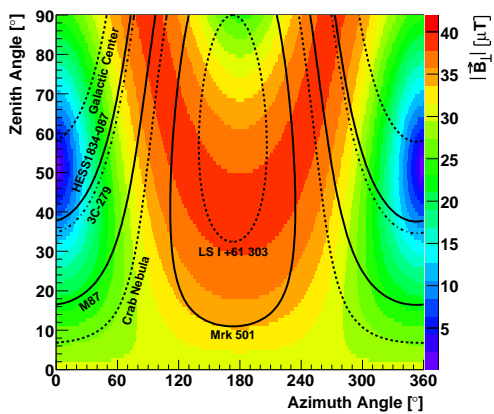}\vspace{-.4cm}
  \caption{The absolute value of the vertical component of the GF strength at the Roque de
    los Muchachos observatory on La Palma.}\label{fig:fieldstrength}
\end{figure}

The influence of the GF on EAS was already qualitatively discussed in 1953 \cite{cocconi1953} 
and later in \cite{chadwick1999,chadwick2000}.
Charged secondary particles in EAS are deflected by the GF which
causes a broadening of the EAS. The east-west separation of electrons and positrons in EAS due to the Lorentz
force can be non negligible compared to the displacement due to Coulomb 
scattering.
The effect on $\gamma$-ray induced EAS is expected to
be more visible than for hadron induced EAS, as their shape is initially more
regular and the scattering angles occurring in nuclear interactions are typically
larger than that produced by the deflection of secondary charged
particles due to the influence of the GF. The Cherenkov images on ground can be affected in a way that the
threshold energy of an IACT increases \cite{bowden1992} as well as its $\gamma$/hadron separation 
capability is expected to be deteriorated.
The goal of the MC studies carried out in this work was to find out about the
impact of the GF on the extraction of the $\gamma$-ray signal from a VHE
$\gamma$-ray source. Figure \ref{fig:fieldstrength} shows the vertical component $|\vec{B}_\perp|$ of the GF strength 
at the site of the MAGIC telescope \cite{barrio1998}
on the Roque de
los Muchachos observatory on La Palma ($28.8^\circ$\,N,$17.9^\circ$\,W)
for 10\,km a.s.l., calculated for November 2006 for the epoch 2005
International Geomagnetic Reference Field (IGRF) model \cite{ngdc}, together with the trajectories of some established and potential VHE $\gamma$-ray sources.
\vspace{-.4cm}

\begin{figure*}
  \subfigure{\hspace{0.2cm}
    \includegraphics[scale=.33]{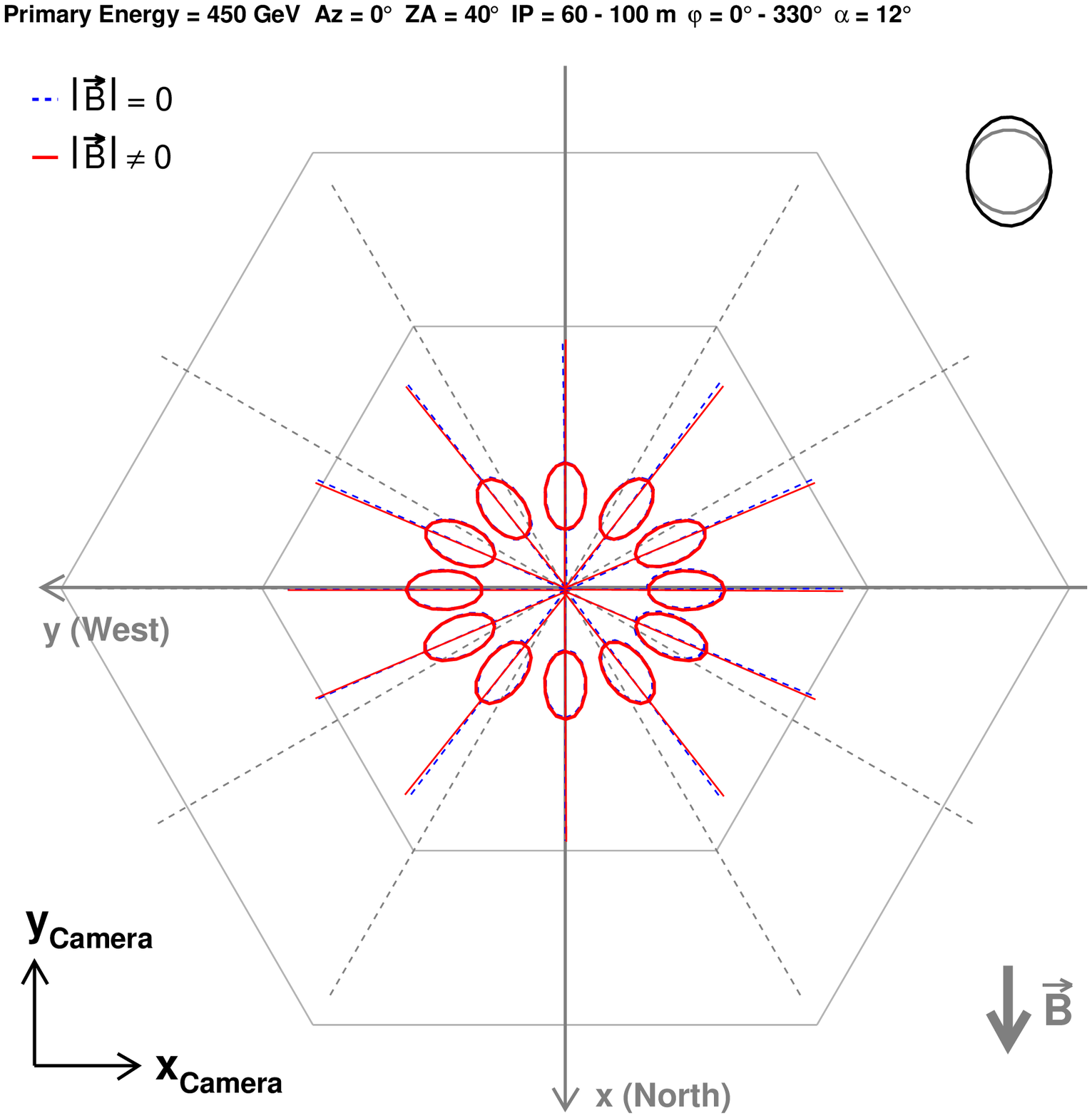}}\qquad\qquad
  \subfigure{
    \includegraphics[scale=.33]{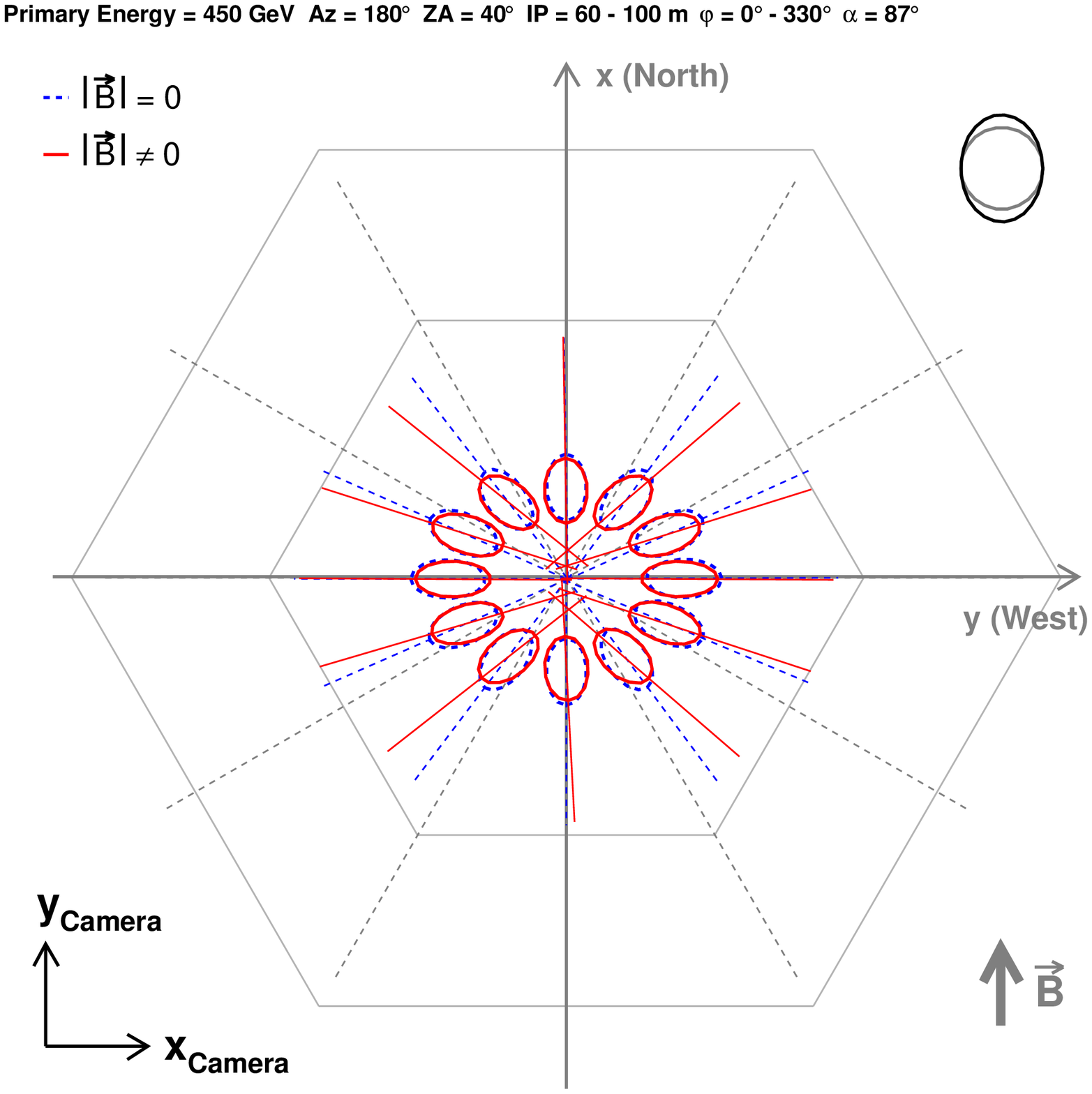}}\vspace{-.1cm}
  \caption{Hillas ellipses in different regions of the camera
  for primary $\gamma$-rays of 450\,GeV energy,
    impact parameters 60\,m\,-\,100\,m, ZA $40^\circ$, azimuth
    angle $0^\circ$ (left), $180^\circ$ (right). Red lines with, blue without
    GF.
  }\label{fig:imageshape}
\end{figure*}

For all sources, the field strength changes noticeably
along the source trajectory. For La Palma, the minimum influence of the GF is expected to occur
in direction of the magnetic north at zenith angle $\text{ZA} =(90^\circ-I)\approx
51^\circ$, where the angle between the shower axis and the GF lines becomes
smallest. $I$ denotes the angle under which the GF lines dip into the Earth's
surface. Hence, the maximum influence is expected to occur for $\text{ZA} = I
\approx 39^\circ$, i.e. for EAS oriented perpendicular to the direction of the
GF lines. It was shown elsewhere \cite{lang1994} that IACT measurements of TeV $\gamma$-rays from the Crab
nebula were not significantly affected when the GF strength was below
$35\,\mu\text{T}$. 

\begin{figure*}
  \subfigure{\hspace{0.5cm}
    \includegraphics[scale=.31]{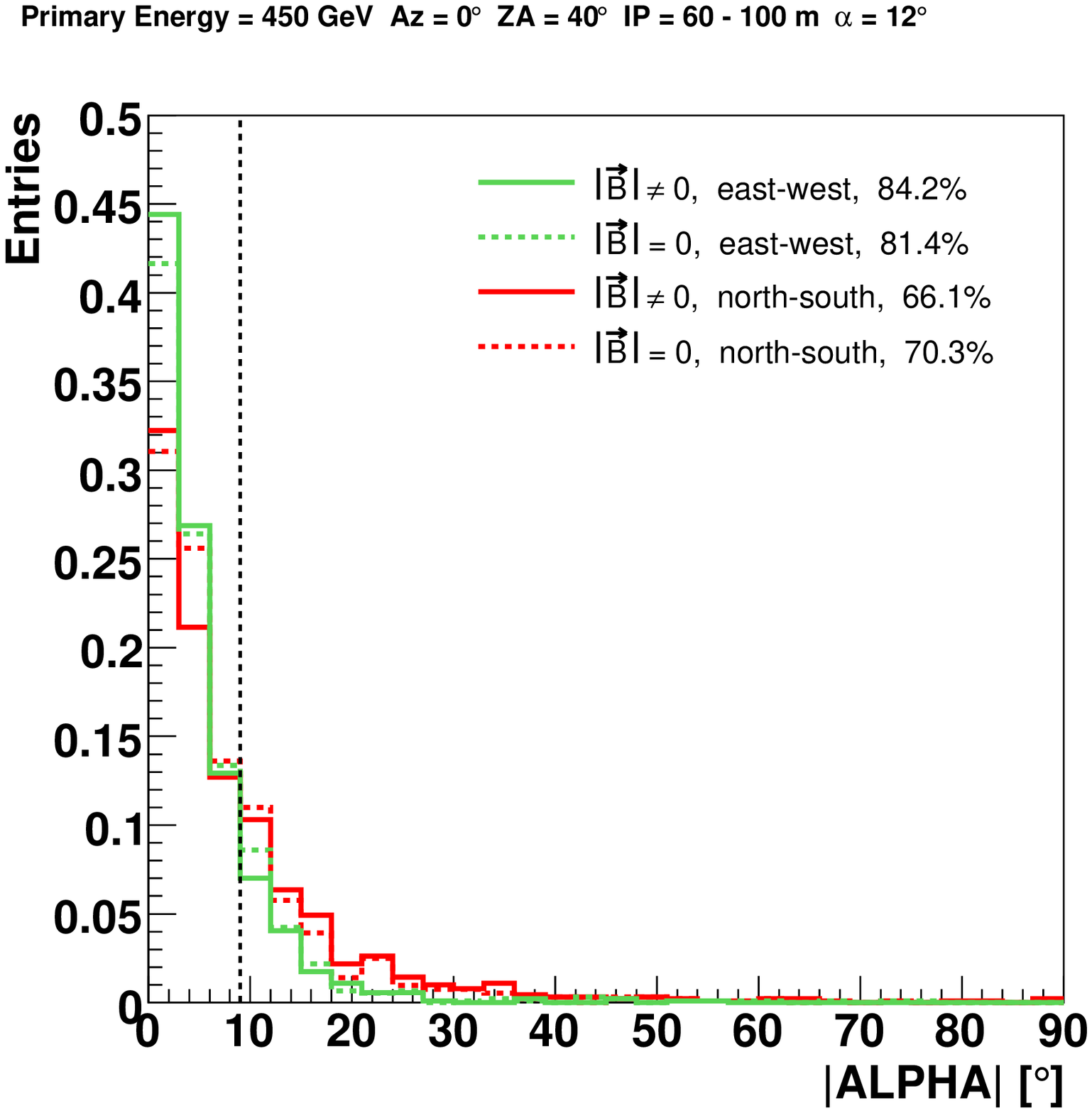}}\qquad\qquad
  \subfigure{
    \includegraphics[scale=.31]{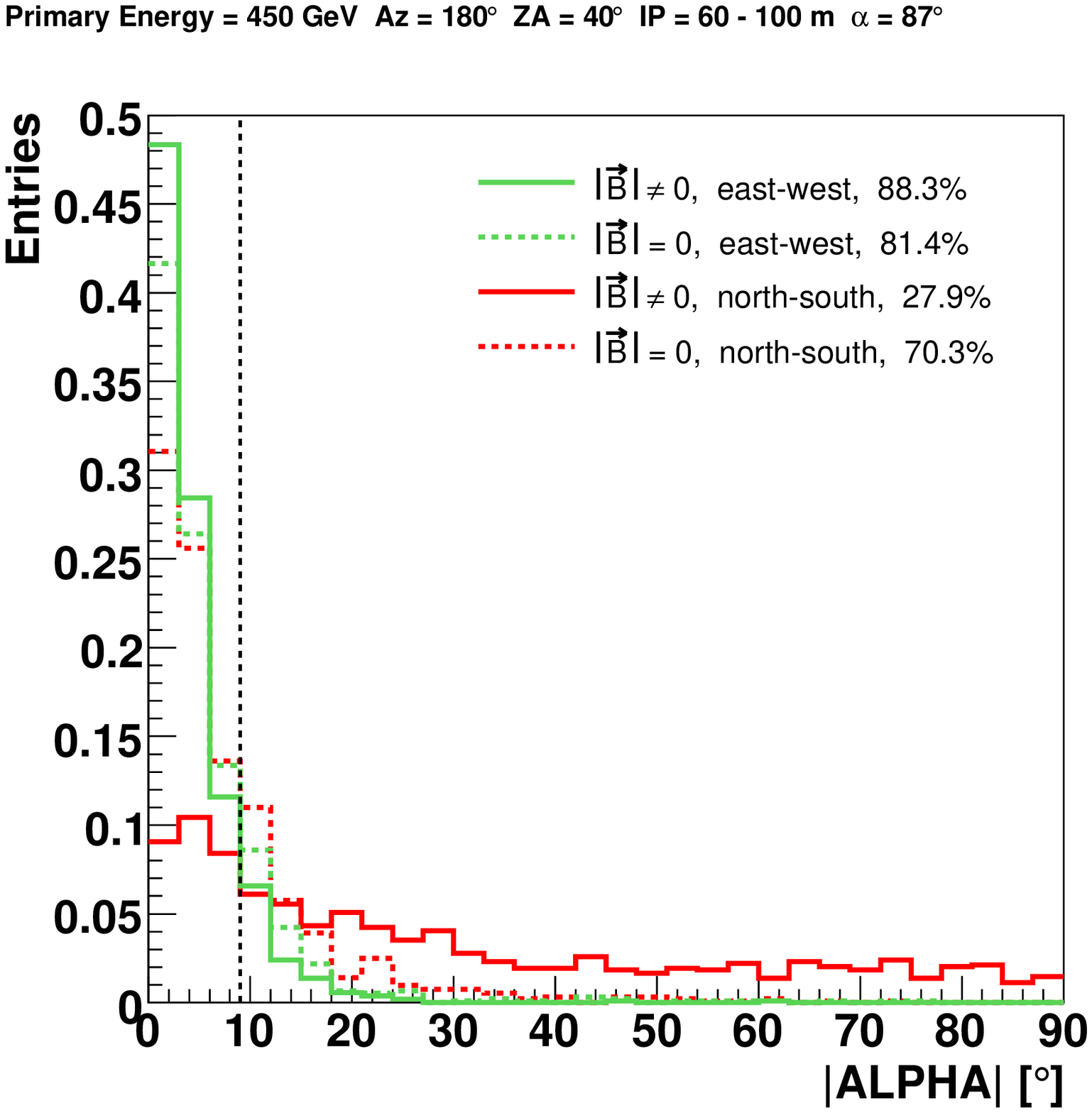}}\vspace{-.5cm}
  \caption{Normalized distributions of the image parameter ALPHA for primary $\gamma$-rays of
    450\,GeV energy, impact parameters  60\,m\,-\,100\,m,
    ZA $40^\circ$, azimuth angle
    $0^\circ$ (left), $180^\circ$ (right).
  }\label{fig:alphaphi}
\end{figure*}

\section{Monte Carlo Simulations \& Analysis}

To study the GF effects, dedicated MC data were produced following
the standard MC production of the MAGIC telescope, doing three
steps \cite{majumdar2005}:

\begin{itemize}
\item[1.] The CORSIKA program (version 6.019) \cite{heck1998} is used to simulate the development of
$\gamma$-ray as well as hadron induced extensive air showers (EAS) for a given
set of input parameters, like the primary $\gamma$-ray energy, the magnitude and direction of
the GF, etc. The GF components were set to the values for La Palma
($28.8^\circ$\,N,$17.9^\circ$\,W) according to the IGRF model \cite{ngdc}. As
a reference, MC data were also produced without GF.

\item[2.] The output of CORSIKA, containing information on the location
and wavelength of each Cherenkov photon on ground, is processed with a dedicated Reflector
program, which does the ray-tracing of the Cherenkov
photons. 

\item[3.] Finally, the output of the Reflector program is
processed by the Camera program simulating the entire readout chain, i.e. 
photomultiplier response, trigger and FADC system including electronic noise.
\end{itemize}

In contrast to the production of standard MC data, where the EAS core
location is randomly placed somewhere in a circle on the plane perpendicular
to the direction of the EAS (to estimate the effective collection area), 
the EAS for this study were simulated 
for fixed impact positions with respect to the
telescope location. This approach allows to investigate the influence of the
GF on the shower images in greater detail.
The calibration and the image parameter calculation (Hillas analysis \cite{hillas1985}) was done using the MAGIC
Analysis and Reconstruction Software (MARS) \cite{bretz2003}.\\
In addition to the MC data 50\,min of low-ZA ($7^\circ$\,-\,$10^\circ$) Crab
nebula data from February 2007 were analyzed considering GF effects.

\section{Results \& Discussion}

Only few selected results can be discussed here and a more detailed analysis
can be found in \cite{seba}.\\
Figure \ref{fig:imageshape} shows the Hillas ellipses for MC simulated
$\gamma$-rays of 450\,GeV energy, $40^\circ$ ZA and impact parameters between
60\,m and 100\,m. The azimuth angle was set to $0^\circ$ (small effect expected, figure
\ref{fig:imageshape}, left image) and $180^\circ$ (strong effect expected, figure
\ref{fig:imageshape}, right image). 
The red ellipses (solid lines) were obtained
for enabled GF in the MC simulation and the blue ones (dashed lines) 
for disabled GF. 
The average orientation is preserved for images oriented either parallel or
vertically with respect to the direction of the GF. Images oriented at
intermediate angles are rotated away from the direction of the GF. 
The extent of the rotation depends on various parameters, like the
$\gamma$-ray energy, the impact parameter, and the position of the EAS with
respect to the telescope.\\
Figure \ref{fig:alphaphi} shows the normalized distributions of the image
parameter ALPHA for MC simulated $\gamma$-rays and the same input parameters
as above.
Again, the azimuth angle was set to $0^\circ$ (small effect expected, figure
\ref{fig:alphaphi}, left image) and $180^\circ$ (strong effect expected, figure \ref{fig:alphaphi}, right
image). 
The ALPHA distributions drawn as red and green solid lines
were obtained for two different directions of the EAS with respect to the
telescope position. However, both distributions correspond to the shower
images that are not rotated (shower images situated on the $x$-axis and
$y$-axis of the CORSIKA coordinate system, figure \ref{fig:imageshape}).
The red distributions correspond to an arrangement where the
connecting line between shower axis and telescope optical axis is parallel to
the north-south direction, whereas the green distributions correspond to an arrangement where the
connecting line between shower axis and telescope optical axis is parallel to
the east-west direction. The ALPHA distributions obtained for disabled GF in the
MC are drawn as red and green dotted lines, respectively. The gray dotted
line indicates the region considered as the signal region.
The MC simulations show that for some arrangements
the ALPHA distribution (red) is significantly degraded
even if the images are not rotated. However, the ALPHA distribution (green) can be slightly enhanced
(stronger peaked at low values) due to the influence of the GF. The remaining possible
arrangements always lead to ALPHA distributions that are degraded due to the
rotation of the shower images.\\
The de-rotation of rotated shower images does not help to recover the pointing
entirely. At most 10\,\% of the events can be recovered by de-rotation of the
shower images. Furthermore, the de-rotation requires the knowledge
of the image parameter.\\ The MC simulations performed for this work also show that the GF significantly affects the energy
reconstruction and the $\gamma$ efficiency.
For unfavorable orientations of
the EAS with regard to the influence of the GF the reconstructed shower image
intensity can be significantly reduced. Therefore, if the GF effects are not taken into account the energy of $\gamma$ candidates from real data will be
systematically underestimated by up to 20\,\%.

\begin{figure}[!h]
  \subfigure{
  \includegraphics[scale= .3]{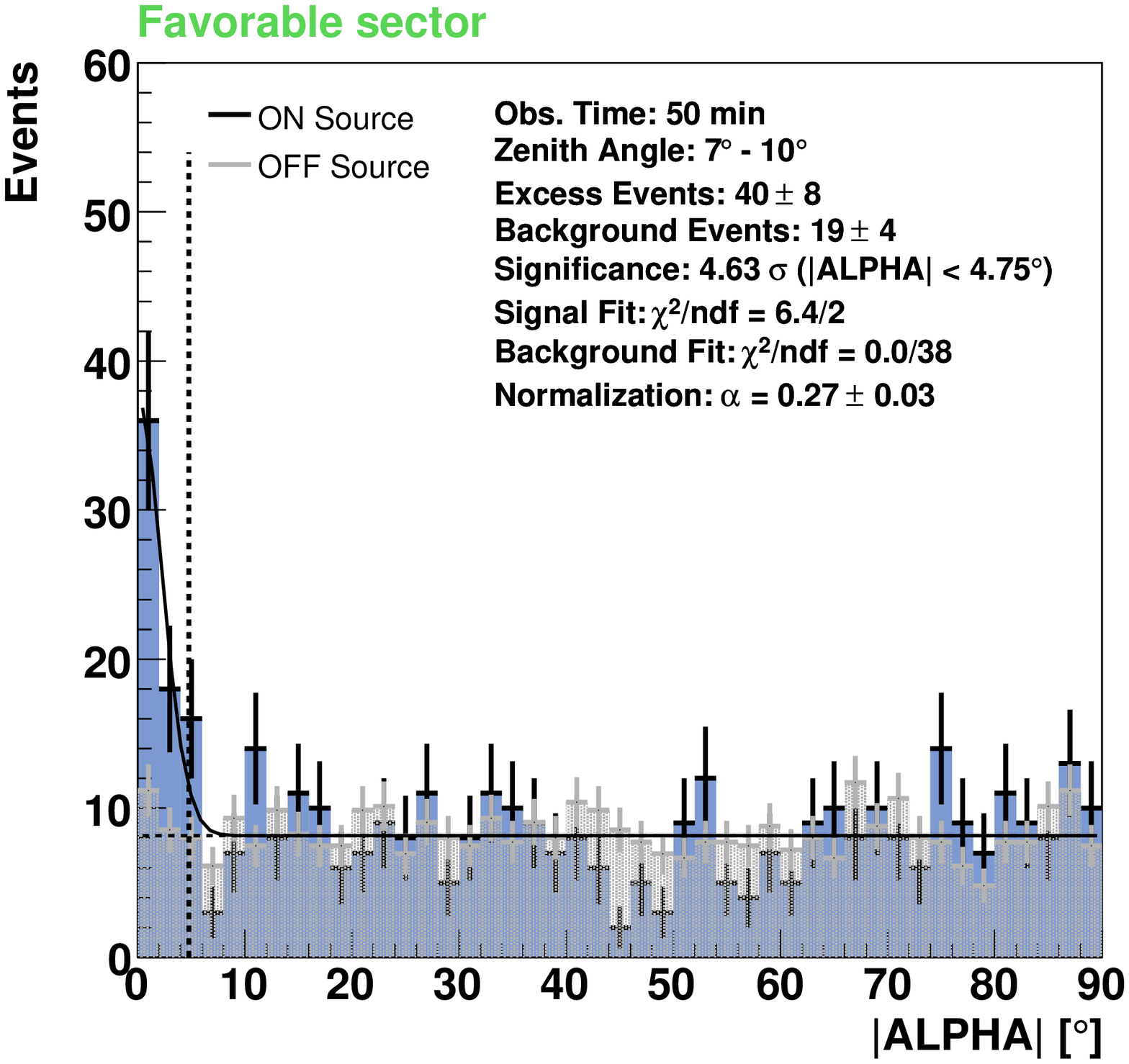}}\\
\subfigure{
  \includegraphics[scale= .3]{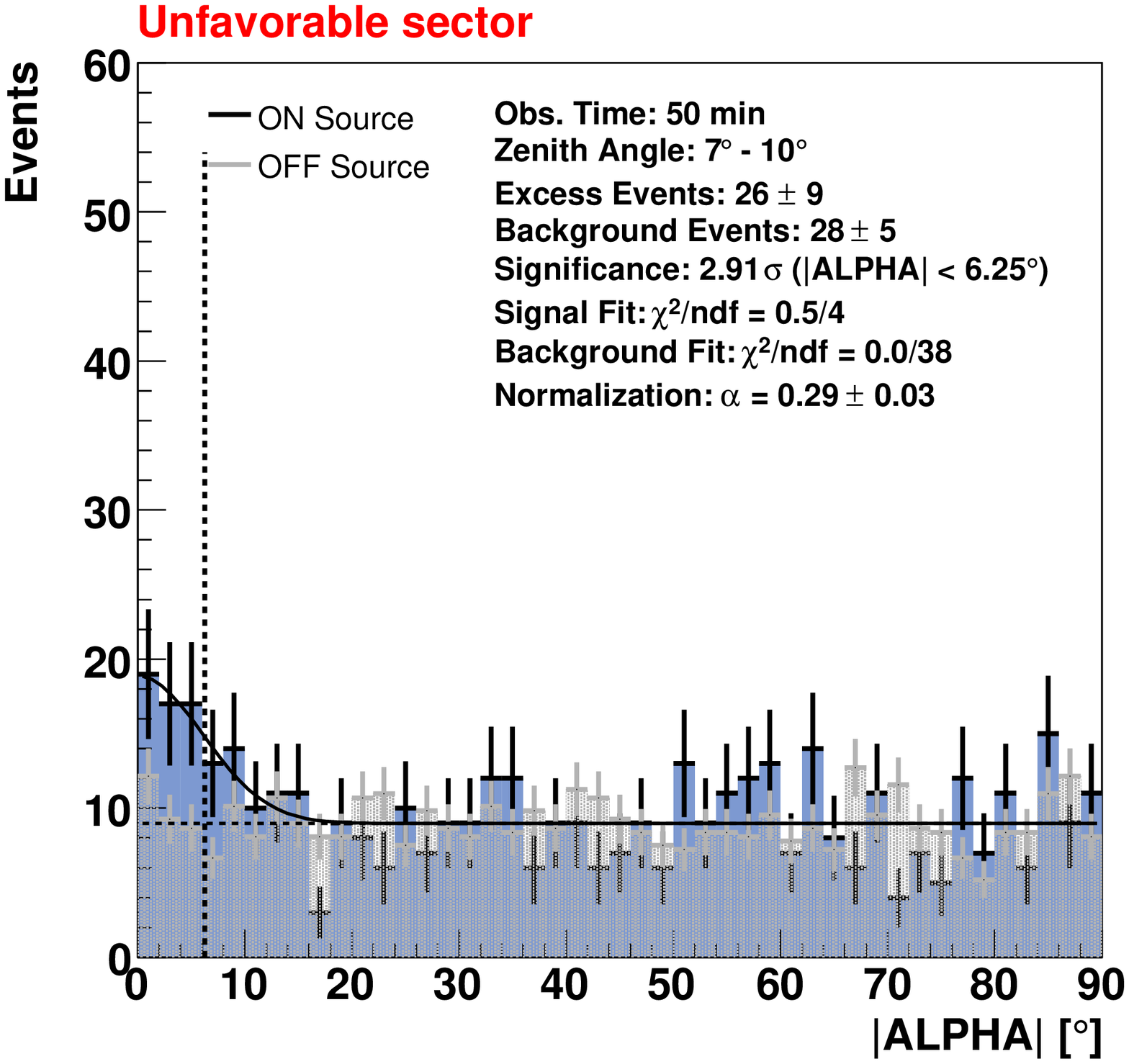}}
\caption{ALPHA distributions for the low-ZA Crab nebula dataset for $\gamma$-ray
  energies above $\sim 120\,\text{GeV}$, considering GF effects.}\label{fig:crabana}
\end{figure}

This is not only the case for low energies but also at higher energies of at least 1\,TeV.
The Cherenkov light distribution on ground from
showers close to the threshold energy can be
thinned out such that most of the events do not survive the trigger level,
i.e. the detection efficiency for $\gamma$-rays can vary by up to 25\,\%
\cite{seba}.\\
Figure \ref{fig:crabana} shows the ALPHA distributions for the small low-ZA Crab nebula dataset for $\gamma$-ray
energies above $\sim 120\,\text{GeV}$, considering GF effects. The upper
distribution corresponds to events oriented at the most favorable camera sector and the lower figure 
to events oriented at the most unfavorable directions with regard to the
influence of the GF, i.e. images are oriented either vertically or
parallel with respect to the direction of the GF. 

\section{Conclusions}

The results from the MC studies suggest that the
influence of the GF can significantly reduce the $\gamma$/hadron separation
capability, the energy estimation and the $\gamma$ efficiency of an
IACT. Altogether, the GF is expected to affect the $\gamma$-ray sensitivity of an
IACT and the determination of both the differential $\gamma$-ray flux and 
the absolute flux level of a $\gamma$-ray source candidate.
Furthermore, the MC studies on the GF effect indicate that appropriate MC 
datasets are not only required for the analysis of low-energy data $\lesssim
100\,\text{GeV}$ but 
also for the reconstruction of VHE $\gamma$-rays of at least 1\,TeV \cite{seba}. The analysis of low-ZA Crab nebula data taken with the
MAGIC telescope proofs that the instrument is sensitive enough to demonstrate
the influence of the GF even for a very low vertical 
component of the GF ($|\vec{B}_{\perp}| < 30\,\mu\text{T}$).

\end{document}